\documentstyle[aps,manuscript,floats,epsf]{revtex}
\tighten
\begin{document}
\def\bq{\begin{equation}}
\def\eq{\end{equation}}
\def\ba{\begin{eqnarray}}
\def\ea{\end{eqnarray}}

\def\hph{\hphantom{-}}
\def\vsk#1{\noalign{\vskip#1 cm}}
\newcommand{\beq}{\begin{equation}}
\newcommand{\eeq}{\end{equation}}
\newcommand{\bea}{\begin{eqnarray}}
\newcommand{\eea}{\end{eqnarray}}

\font\fortssbx=cmssbx10 scaled \magstep2
\hbox to \hsize{
{\fortssbx University of Wisconsin - Madison}
\hfill$\vtop{\hbox{\bf MADPH-97-999}
                \hbox{\bf DOE-ER40757-100}
                \hbox{\bf KEK-TH-529}
                \hbox{\bf hep-ph/9707412}
                \hbox{July 1997}}$ }

\vspace{.75in}

\begin{center}
{\large\bf Global Study of Electron-Quark Contact Interactions }

V. Barger$^1$, Kingman Cheung$^2$, K. Hagiwara$^{1,3,4}$,
and D. Zeppenfeld$^1$\\[5mm]
\it
$^1$Physics Department, University of Wisconsin, Madison, WI 53706\\
$^2$Center for Particle Physics, University of Texas, Austin, TX 78712\\
$^3$Theory Group, KEK, Tsukuba, Ibaraki 305, Japan\\
$^4$ICEPP, University of Tokyo, Hongo, Bunkyo-ku, Tokyo 113, Japan
\end{center}

\begin{abstract}
We perform a global fit of data relevant to $eeqq$ contact
interactions, including deep inelastic scattering at high $Q^2$
from ZEUS and H1, atomic physics parity violation in Cesium from JILA,
polarized $e^-$ on nuclei scattering experiments at SLAC, Mainz and Bates,
Drell-Yan production at the
Tevatron, the total hadronic cross section $\sigma_{\rm had}$ at LEP,
and neutrino-nucleon scattering
from CCFR.
With only the new HERA data, the presence of contact interactions
improves the fit compared to the Standard Model. When other data sets
are included, the size of the contact contributions is reduced and
the overall fit represents no real improvement over the Standard Model.
\end{abstract}

\thispagestyle{empty}
\newpage

\section{Introduction}

The reports of event rates above Standard Model (SM) expectations
in $e^+p\to e^+X$ deep inelastic scattering (DIS)
at very high $Q^2$ by the H1~\cite{H1}
and ZEUS~\cite{zeus} experiments at HERA have generated a considerable amount
of theoretical activity. Various models to explain the excess events have been
put forward\cite{alta,babu,lq,kon91,squark,comp,schnitzer84,ours,nelson,%
barto,buch,desh,contact,kuhlmann}.
Particular attention has been given to resonant production
of leptoquarks\cite{alta,babu,lq}, squarks with $R$-parity violating
couplings\cite{kon91,squark}, quark and lepton compositeness\cite{comp},
and a general parametrization in terms of $eeqq$ contact
interactions\cite{alta,babu,schnitzer84,ours,nelson,barto,buch,desh,contact},
representing the exchange of very massive particles in the $s$, $t$
or $u$-channels.

An interest in leptoquarks derives from an indication for an $e^+j$
invariant mass peak at $M_{ej}\approx 200$~GeV in the H1 data.
However, no clear mass peak is present in the ZEUS data; see discussion in
Ref.~\cite{drees97}.
In addition, for an $ej$ branching
fraction of unity, recent searches at the Tevatron exclude leptoquarks
of mass $M_{LQ}\leq 210$~GeV from CDF~\cite{lqCDF} and
$M_{LQ}\leq 225$~GeV from D0~\cite{lqD0}. The Tevatron bounds can
be escaped, however, with a reduced branching fraction $B(LQ\to ej)$.

Any new physics in $eq\to eq$ scattering for which the exchanged
particles have mass squared $M^2\gg s$ can be described by an
effective $eeqq$ contact term Lagrangian. For example, effects of a $Z'$ boson
of TeV
mass scale would be well represented by a four-fermion contact interaction.
A leptoquark of mass $M_{LQ}\gg 200$~GeV could similarly be described
at HERA by an effective $eeqq$ contact term. For leptoquark masses not far
above the HERA energy range additional form-factor effects would have to be
considered, but an analysis in terms of contact interactions would still
be appropriate to derive constraints from low energy experiments.

In this paper we perform a combined  analysis of low and high energy data
to obtain constraints on $eeqq$ contact interactions. The data sets include
deep inelastic scattering at high $Q^2$ from ZEUS~\cite{zeus} and
H1~\cite{H1}, atomic physics parity violation~\cite{apv}, polarized
$e^-$ on nuclei scattering experiments at SLAC~\cite{eD}, Mainz\cite{mainz89},
and Bates\cite{bates90}, Drell-Yan production at the
Tevatron~\cite{cdf-dy}, the total hadronic cross section $\sigma_{\rm had}$
at LEP1, LEP1.5, and LEP2~\cite{lep,opal,L3,aleph,langacker2},
the left-right asymmetry $A_{LR}$ at SLD~\cite{lep}, and
neutrino-nucleon scattering from CCFR~\cite{ccfr}.

The organization of the paper is as follows. In Section~II we discuss
the parametrization of contact interactions and discuss possible symmetry
relations among them. In Section~III we present all the data that enter our
global analysis. In Section~IV we give constraints on the $eeqq$ contact
interactions in selected models and show how each experiment constrains
different combinations of these interactions. Allowing for all eight $eeuu$ and
$eedd$ contact interactions to vary freely, the high $Q^2$ HERA data and the
high mass Drell-Yan pair data from CDF cannot be fitted simultaneously.
Section~V gives our conclusions.

\section{Parametrization of Contact Interactions}

The conventional effective Lagrangian of $e e q q$ contact
interactions has the form~\cite{ELP,cashmore,chiap}
\begin{eqnarray}
L_{NC} &=& \sum_q \Bigl[ \eta_{LL}
\left(\overline{e_L} \gamma_\mu e_L\right)
\left(\overline{q_L} \gamma^\mu q_L \right)
+ \eta_{RR} \left(\overline{e_R}\gamma_\mu e_R\right)
                 \left(\overline{q_R}\gamma^\mu q_R\right) \nonumber\\
&& \qquad {}+ \eta_{LR} \left(\overline{e_L} \gamma_\mu e_L\right)
                             \left(\overline{q_R}\gamma^\mu q_R\right)
+ \eta_{RL} \left(\overline{e_R} \gamma_\mu e_R\right)
\left(\overline{q_L} \gamma^\mu q_L \right) \Bigr] \,, \label{effL}
\end{eqnarray}
where the eight independent coefficients $\eta_{\alpha\beta}^{eu}$ and
$\eta_{\alpha\beta}^{ed}$ have dimension (TeV)$^{-2}$ and are conventionally
expressed as $\eta_{\alpha\beta}^{eq} = \epsilon g^2 /\Lambda_{eq}^2$,
with a fixed $g^2=4\pi$.
The sign factor $\epsilon= \pm 1$ allows for either constructive or destructive
interference with the SM $\gamma$ and $Z$ exchange amplitudes and
$\Lambda_{eq}$ represents the mass scale of the exchanged new
particles, with coupling strength $g^2/4\pi=1$. A coupling of this
order is expected in substructure models and therefore $\Lambda_{eq}$ is
sometimes called the ``compositeness scale''~\cite{ELP}.
In the effective interaction (\ref{effL}) we do not
include lepton or quark chirality violating terms  such as
$\left(\overline{e_L} e_R\right) \left(\overline{q_L} q_R\right)$ since, if
there is an approximate
$SU(2)\times U(1)$ invariance,
scalar and tensor contact terms are severely constrained by helicity
suppressed processes like $\pi^-\to e^-\bar\nu$~\cite{alta}. A contact
term Lagrangian completely analogous to Eq.~(\ref{effL}) can be written
down for $\nu\nu qq$ interactions, with parameters $\eta^{\nu q}_{LL}$
and $\eta^{\nu q}_{LR}$, respectively.

\subsection{$SU(2)\times U(1)$ symmetry and universality }

Left-handed electrons and quarks belong to SU(2) doublets $L=(\nu_L,e_L)$
and $Q=(u_L,d_L)$ and thus from SU(2) symmetry
one expects relations between contact terms involving left-handed
$u$ or $d$ quarks; similarly, contact terms for left-handed electrons and
neutrinos should be related\cite{desh}. In order to exhibit these relationships
we start from the most general $SU(2)\times U(1)$ invariant contact term
Lagrangian,
\begin{eqnarray}
{\cal L}_{SU(2)}&=&
\eta_1 \Bigl(\overline L\gamma^\mu L\Bigr) \Bigl(\overline Q\gamma_\mu Q\Bigr)
+
\eta_2 \Bigl(\overline L\gamma^\mu T^aL\Bigr) \Bigl(\overline Q\gamma_\mu T^a
Q\Bigr) + \eta_3 \Bigl(\overline L\gamma^\mu L\Bigr)
\Bigl(\overline{u_R}\gamma_\mu
u_R\Bigr)\nonumber \\
&& {}+ \eta_4 \Bigl(\overline L\gamma^\mu L\Bigr)
\Bigl(\overline{d_R}\gamma_\mu
d_R\Bigr)
+ \eta_5 \Bigl(\overline{e_R}\gamma^\mu e_R\Bigr) \Bigl(\overline Q\gamma_\mu
Q\Bigr) +
\eta_6 \Bigl(\overline{e_R}\gamma^\mu e_R\Bigr) \Bigl(\overline{u_R}\gamma_\mu
u_R\Bigr) \nonumber\\
&& {}+ \eta_7 \Bigl(\overline{e_R}\gamma^\mu e_R\Bigr)
\Bigl(\overline{d_R}\gamma_\mu
d_R\Bigr)\; .
\label{effLsu2}
\end{eqnarray}
Here we have suppressed the lepton chirality violating term
$\left(\overline L\gamma^\mu Q\right) \left(\overline{d_R}\gamma_\mu
e_R\right)$
which, under a Fierz
transformation, is equivalent to a $\left(\overline L
e_R\right)\left(\overline{d_R}Q\right)$
term which, as noted before, is severely constrained by $\pi^-\to e^-\bar\nu$
data. In the second term of Eq.~(\ref{effLsu2})
$T^a=\sigma^a/2$ denotes the SU(2) generators; this term describes
the exchange of isospin triplet quanta between the lepton and quark fields.
Because of this possible contribution no constraints arise for the
$\eta^{eq}_{LL}$
contact terms and the difference $\eta^{ed}_{LL}-\eta^{eu}_{LL}=\eta_2/2$
measures the size of $e\nu ud$ contact terms in CC processes.
SU(2) does constrain right-handed electron couplings,
\begin{equation}
\eta^{eu}_{RL}=\eta_5=\eta^{ed}_{RL} \; .
\label{su2releR}
\end{equation}
In addition, the four neutrino and the lepton couplings are related by SU(2),
\begin{eqnarray}
\eta^{\nu u}_{LL} & = &\eta_1\;+\;{1\over 4}\eta_2 = \eta^{ed}_{LL}\; ,
\nonumber \\
\eta^{\nu d}_{LL} & = &\eta_1\;-\;{1\over 4}\eta_2 = \eta^{eu}_{LL}\; ,
\nonumber \\
\eta^{\nu u}_{LR} & = &\eta_3 = \eta^{eu}_{LR}\; ,
\nonumber \\
\eta^{\nu d}_{LR} & = &\eta_4 = \eta^{ed}_{LR}\; .
\label{su2relnu}
\end{eqnarray}

In our analyses, the relations of Eqs.~(\ref{su2releR}) and (\ref{su2relnu})
are only used when neutrino scattering data are included in the analysis. Even
though we expect that $SU(2)\times U(1)$ will be a
symmetry of the renormalizable interactions which ultimately manifest
themselves
as the contact terms of Eq.~(\ref{effL}), electroweak symmetry breaking may
break the degeneracy of SU(2) multiplets of new, heavy quanta whose exchanges
give rise to (\ref{effL}). This would result in a violation of the relations
of Eqs.~(\ref{su2releR}) and (\ref{su2relnu}). One example is the exchange
of the stop $\tilde t_1$, $\tilde t_2$, and the sbottom $\tilde b_L, \tilde
b_R$
in $R$-parity violating SUSY models. The large top-quark mass may lead to
substantial splitting of the masses of these squarks which could easily
lead to violations by up to a factor of two of SU(2) relations such as
$\eta^{\nu d}_{LR}=\eta^{ed}_{LR}$.

In the discussion above we have considered first generation quarks and leptons
only, because the ``HERA anomaly'' raises particular interest in such
couplings. In principle, all $\eta$'s carry four independent generation
indices and may give rise to other flavor conserving and flavor changing
transitions. Because of severe
experimental constraints on intragenerational transitions~\cite{wyler} like
$K\to\mu e$
we restrict our discussion to first generation contact terms. Only
where required by particular data (e.g.\ the LEP data which sum over the quarks
of
all three generations) will we assume universality of contact terms between
$e$, $\mu$ and $\tau$ and up- and down-type quarks of different generations.

Other symmetry constraints on the parametrization of four-fermion contact
interactions are of interest as well:
\begin{enumerate}
\item[i)]  SU(12) symmetry considerations\cite{nelson} give
\begin{equation}
\eta^{eq}_{\alpha L} =- \eta^{eq}_{\alpha R}\;, \qquad \alpha=L\;\;{\rm or}
\;\; R \;.
\end{equation}
In addition, SU$(2)_L$ gauge invariance requires the $u$-quark
contribution equal to the $d$-quark contribution, i.e.,
\begin{equation}
\eta^{eu}_{\alpha L} = \eta^{ed}_{\alpha L}\;, \qquad  \alpha= L\;\;
{\rm or}\;\; R  \;.
\end{equation}
With these symmetries the atomic parity violation constraint is satisfied
naturally.
\item[ii)] Vector-vector (VV) interactions \cite{barto}, from the exchange of a
boson with purely vector couplings, give
\begin{equation}
\eta^{eq}_{LL}=\eta^{eq}_{RR}=\eta^{eq}_{LR}=\eta^{eq}_{RL}\equiv
\eta^{eq}_{VV} \;.
\end{equation}

\item[iii)] Axial-axial interactions \cite{barto,buch}, from the exchange of
purely axial-vector couplings, give
\begin{equation}
\eta^{eq}_{LL}=\eta^{eq}_{RR}=-\eta^{eq}_{LR}=-\eta^{eq}_{RL}\equiv
\eta^{eq}_{AA} \;.
\end{equation}
\end{enumerate}
We consider fits with these restrictions imposed subsequently to make
model-independent analysis.

\subsection{Observables and four-fermion amplitudes}

All the observables that we consider are described by a four-fermion
$S$-matrix element. The amplitudes for observables
such as $e^+e^-\to $~hadrons, $p\bar p \to \ell^+\ell^-X$ and atomic parity
violation,
are obtained from the amplitude for $eq\to eq$ scattering by crossing;
they are given by angular factors times reduced
amplitudes $M^{eq}_{\alpha \beta}$, where the subscripts label the chiralities
of the initial lepton ($\alpha$) and quark ($\beta$). The SM tree level reduced
amplitude for $eq\to eq$ is
\begin{equation}
\label{reduced}
M^{eq}_{\alpha\beta}(\hat t) = -\frac{e^2 Q_q}{\hat t} +
\frac{e^2 }{\sin^2 \theta_{\rm w}  \cos^2 \theta_{\rm w}}
\frac{g_\alpha^e g_\beta^q}{\hat t - m_Z^2} \;,
\qquad \alpha,\beta = L,R
\end{equation}
where $\hat t = -Q^2$ is the Mandelstam variable,
$g_L^f = T_{3f} - \sin^2 \theta_{\rm w} Q_f$ and
$g_R^f = -\sin^2 \theta_{\rm w} Q_f$, $T_{3f}$ and $Q_f$ are, respectively,
the third component of the SU(2) isospin and the electric charge of the
fermion $f$ in units of the proton charge, and $e^2=4\pi \alpha_{\rm em}$.
For $e^+e^-\to q\bar q$ or $q\bar q\to e^+e^-$, the amplitude
$M_{\alpha\beta}^{eq}(\hat s)$ is obtained from Eq.~(\ref{reduced}) with the
replacement $\hat t\to \hat s$ and $\hat t-M_Z^2 \to \hat s-M_Z^2+i\hat s
\Gamma_Z/M_Z$,  where $\hat s$ is the subprocess c.m.\ energy squared.
The new physics contributions to the reduced amplitudes $M_{\alpha\beta}$ from
the $eeqq$ contact interactions of Eq.~(\ref{effL}) are
\begin{equation}
\Delta M^{eq}_{\alpha \beta}  = \eta^{eq}_{\alpha \beta}\;, \qquad
\alpha,\beta = L,R \;. \label{newphys}
\end{equation}

\subsection{Contributions to Standard Model parameters}

In low energy neutral current (NC) processes, at $\sqrt{s}\ll m_Z^2$,
$Z$ boson exchange can also be described by effective four-fermion
contact terms. For the parity violating contributions to $eeqq$ NC
interactions the NC Lagrangian is conventionally expressed in terms
of parameters $C_{1q}$ and $C_{2q}$ as
\begin{equation}
\label{lag}
{\cal L}^{e \,{\rm Hadron}} = \frac{G_F}{\sqrt{2}} \; \sum_q \biggl[
C_{1q} \left(\bar e \gamma^\mu \gamma^5 e\right) \; \left(\bar q \gamma_\mu
q\right) + C_{2q} \left(\bar e \gamma^\mu  e \right)\; \left(\bar q \gamma_\mu
\gamma^5 q\right) \biggr] \;.
\end{equation}
The radiatively corrected SM values are~\cite{pdg}
\begin{eqnarray}
C_{1q}^{\rm SM} &=&
\rho'_{eq}\left[-T_{3q} +2Q_q (\kappa'_{eq} \sin^2\theta_{\rm w})\right]\; ,
\nonumber \\
C_{2q}^{\rm SM} &=&
-T_{3q}\;\rho_{eq}\left[1-4\; (\kappa_{eq} \sin^2\theta_{\rm w})\right]\;
+\; \lambda_{2q}\; ,
\label{c1}
\end{eqnarray}
where $\sin^2\theta_{\rm w} = 0.2236$, $\rho'_{eq}=0.9884$
and $\kappa'_{eq}=1.036$ for the atomic parity violation experiments
while $\rho'_{eq}=0.979$, $\kappa'_{eq}=1.034$, $\rho_{eq}=1.002$,
$\kappa_{eq}=1.06$, $\lambda_{2u}=-0.013$, and $\lambda_{2d}=0.003$ for the
lepton-nucleon scattering.
The contact term Lagrangian introduces parameter shifts
\begin{eqnarray}
\Delta C_{1q} &=& \frac{1}{2\sqrt{2} G_F} \biggr[
-  \eta_{LL}^{eq} +\eta_{RR}^{eq} -\eta_{LR}^{eq} +\eta_{RL}^{eq} \biggr] \;,
\nonumber \\
\Delta C_{2q} &=& \frac{1}{2\sqrt{2} G_F} \biggr[
-  \eta_{LL}^{eq} +\eta_{RR}^{eq} +\eta_{LR}^{eq} -\eta_{RL}^{eq} \biggr]\;,
\qquad q = u,d
\end{eqnarray}
that can be constrained by experimental data on atomic parity violation and
electron nucleon scattering.

If $SU(2)\times U(1)$ is a good symmetry of the contact interactions then
neutrino-nucleon scattering data also constrain the $eeqq$ contact
terms~\cite{desh}. Comparing to the conventional parametrization of
neutrino-quark effective interactions,
\begin{equation}
{\cal L} = -\, \frac{G_F}{\sqrt{2}}\; \bar \nu \gamma^\mu (1-\gamma_5) \nu
\; \sum_q \biggr [
g_L^q \;\bar q \gamma_\mu (1-\gamma_5) q +
g_R^q \;\bar q \gamma_\mu (1+\gamma_5) q
\biggr ] \;, \label{effLnuq}
\end{equation}
the contact interactions (\ref{effL}) introduce shifts in the coefficients
$g_{L,R}^q$,
\begin{equation}
\Delta g_L^q =  - \frac{1}{2\sqrt{2}G_F} \eta_{LL}^{eq}\;,\qquad
\Delta g_R^q =  - \frac{1}{2\sqrt{2}G_F} \eta_{LR}^{eq}\;,
\qquad q=u,d \;. \label{shift}
\end{equation}
Here we have used the $SU(2)\times U(1)$ relations of Eq.~(\ref{su2relnu}).
The SM values for these couplings are\cite{pdg}
%
\begin{eqnarray}
\left(g_L^q\right)^{\rm SM} &=& 1.0095\left(T_{3q} -
Q_q\;(1.0382\sin^2\theta_{\rm w}) +
\lambda_{qL}\right) \\
\left(g_R^q\right)^{\rm SM} &=& 1.0095\left(-Q_q\;(1.0382\sin^2\theta_{\rm w})
+  \lambda_{qR}\right) \; ,
\end{eqnarray}
with $\lambda_{uL}=-0.0032$, $\lambda_{dL}=-0.0026$ and
$\lambda_{uR}= \lambda_{dR}/2 = 3.6\times 10^{-5}$.

\subsection{Connection to specific models}

A second $Z$-boson, $Z_2$, of mass $M_{Z_2}$ and with chiral couplings
$e g_\alpha^{f(2)}$ to fermion species $f$, would give rise to four-fermion
contact terms with
\begin{equation}
\eta_{\alpha\beta}^{eq} =
- e^2 \frac{g_\alpha^{e(2)} g_\beta^{q(2)} }{M_{Z_2}^2} \;,
\end{equation}
in experiments at $\hat s\ll M_{Z_2}^2$. For subprocess energies of the
order of the $Z_2$ mass, as may be achievable at the Tevatron, form-factor
effects, due to the $\hat s$ dependence of the $Z_2$ propagator, would have
to be considered.

The exchange of scalar leptoquarks would give rise to contact terms
\begin{equation}
\eta_{\alpha\beta}^{eu} = e^2\left(
 \frac{h_{2L}^2 \delta_{\alpha L} \delta_{\beta R}}{2 (- M_{R_{2L}}^2)}
+\frac{h_{2R}^2 \delta_{\alpha R} \delta_{\beta L}}{2 (- M_{R_{2R}}^2)}
            \right )\;, \qquad
\eta_{\alpha\beta}^{ed} = e^2\left(
 \frac{\tilde{h}_{2L}^2 \delta_{\alpha L} \delta_{\beta R}}
{2 (- M_{\tilde{R}_{2L}}^2)}
+\frac{h_{2R}^2 \delta_{\alpha R} \delta_{\beta L}}{2 (- M_{R_{2R}}^2)}
\right) \;.
\end{equation}
where $eh_{2L}, eh_{2R}, e\tilde{h}_{2L}$ are the couplings of
$R_{2L}, R_{2R}, \tilde{R}_{2L}$, respectively \cite{brw}.
A leptoquark mass in the 200--300~GeV range would again necessitate the
inclusion of form-factor (propagator) effects in the analysis of HERA and
Tevatron data. For the low-energy data, however, the contact term
parametrization would be entirely adequate.

\section{Experimental Data Sets}

\subsection{ZEUS and H1: $e^+ p \to e^+ X$}

The H1 and ZEUS experiments have provided event rate measurements in $x,y$
bins where $x$ and $y$ are the usual scaling variables of DIS:
\begin{equation}
x = Q^2/ \left[ 2p\cdot (k-k') \right], \qquad y = Q^2 / [sx] \,.
\end{equation}
Here $k$ and $k'$ are the four-momenta of the incoming electron and outgoing
positron, respectively, $p$ is the proton four-momentum, $s\simeq(300\rm\
GeV)^2$ is
the square of the c.m.\ energy, and $Q^2$ is the square of the momentum
transfer
\begin{equation}
Q^2 = -(k-k')^2 = sxy \,.
\end{equation}
The observed distribution of events in $x,y$ bins is given in Table~I for
ZEUS\cite{zeus} and in Table~II for H1\cite{H1}, along
with SM expectations that include experimental efficiencies.
Systematic errors are correlated between bins. However, these correlations
have not been published yet and we assume their effects to be small
compared to the large statistical errors of the published data.

Neutral current deep inelastic scattering occurs via the subprocess $eq\to eq$.
In terms of the reduced amplitudes
$M^{eq}_{\alpha \beta} (\hat t) \;(\alpha,\beta=L,R)$ of Eq.~(\ref{reduced})
the spin- and color-averaged amplitude-squared for
$e^+(k) q (p) \to e^+ (k') q(p')$  is
given by
\begin{equation}
\overline{\sum} |{\cal M}(e^+ q\to e^+ q)|^2 =
 \biggr( |M^{eq}_{LL}(\hat t)|^2 + |M^{eq}_{RR}(\hat t)|^2 \biggr) \hat u^2 +
 \biggr( |M^{eq}_{LR}(\hat t)|^2 + |M^{eq}_{RL}(\hat t)|^2 \biggr) \hat s^2
   \;, \label{ampl^2}
\end{equation}
where $\hat t = -sxy$, $\hat u = -sx(1-y)$ and $\hat s = sx$.
In our analysis we calculate the SM tree level cross section for each $x,y$
bin. We then compare with the efficiency corrected SM event rates,
$N^{\rm SM}(x,y; \mbox{corrected})$, in Tables~I and II
to obtain correction factors for each bin,
\begin{equation}
C(x,y) = {N^{\rm SM}(x,y; \mbox{corrected})\over \sigma^{\rm SM}(x,y)}\; .
\end{equation}

Then in analyzing models that include new physics contributions we multiply the
theoretical cross sections by $C(x,y)$ to take into account the experimental
efficiencies and compare the corrected result with the observed number of 
events.
We treat the H1 and ZEUS data separately.
The SM DIS differential cross section is given by
\begin{eqnarray}
d\sigma(e^+p)\over dx\,dy &=& {sx\over16\pi} \left\{ u(x,Q^2) \left[
\left| M_{LR}^{eu}(\hat t) \right|^2 + \left| M_{RL}^{eu}(\hat t) \right|^2
+ (1-y)^2 \left(
\left| M_{LL}^{eu}(\hat t) \right|^2 + \left| M_{RR}^{eu}(\hat t) \right|^2
\right) \right] \right.  \nonumber\\
&&  {}+ d(x,Q^2) \left[ \left| M_{LR}^{ed} (\hat t)\right|^2
+ \left| M_{RL}^{ed}(\hat t) \right|^2 + (1-y)^2
\left( \left| M_{LL}^{ed}(\hat t) \right|^2 + \left|
M_{RR}^{ed}(\hat t) \right|^2 \right) \right]
\nonumber \\
&&+\bar u(x,Q^2)\biggr[\left|M_{LL}^{eu}(\hat t) \right|^2
            +\left |M_{RR}^{eu}(\hat t)\right|^2 +
 (1-y)^2  \left(\left|M_{LR}^{eu}(\hat t)\right|^2 +
    \left|M_{RL}^{eu}(\hat t)\right|^2
	            \right)  \biggr ]
\nonumber \\
&& + \bar d(x,Q^2)\biggr[ \left|M_{LL}^{ed}(\hat t)\right|^2
+ \left|M_{RR}^{ed}(\hat t)\right|^2 +
          (1-y)^2  \left( \left|M_{LR}^{ed}(\hat t)\right|^2
 + \left|M_{RL}^{ed}(\hat t)\right|^2
	            \right)  \biggr ]
\nonumber \\
&& + s(x,Q^2) \biggr[ \left|M_{LR}^{es}(\hat t)\right|^2 +
  \left|M_{RL}^{es}(\hat t)\right|^2 +
      (1-y)^2 \left ( \left|M_{LL}^{es}(\hat t)\right|^2
   + \left|M_{RR}^{es}(\hat t)\right|^2
		\right ) \biggr ]
\nonumber \\
&& + \bar s(x,Q^2) \biggr[ \left|M_{LL}^{es}(\hat t)\right|^2 +
  \left|M_{RR}^{es}(\hat t)\right|^2 +
             (1-y)^2 \left( \left|M_{LR}^{es}(\hat t)\right|^2 +
  \left|M_{RL}^{es}(\hat t)\right|^2
                 \right) \biggr]
\nonumber\\
&& + c(x,Q^2) \biggr[ \left|M_{LR}^{ec}(\hat t)\right|^2 +
 \left|M_{RL}^{ec}(\hat t)\right|^2 +
      (1-y)^2 \left ( \left|M_{LL}^{ec}(\hat t)\right|^2 +
 \left|M_{RR}^{ec}(\hat t)\right|^2
		\right ) \biggr ]
\nonumber \\
&&  \left.  + \bar c(x,Q^2) \biggr[ \left|M_{LL}^{ec}(\hat t)\right|^2 +
  \left|M_{RR}^{ec}(\hat t)\right|^2 +
             (1-y)^2 \left( \left|M_{LR}^{ec}(\hat t)\right|^2 +
 	\left|M_{RL}^{ec}(\hat t)\right|^2
                   \right) \biggr]
\right\}  \,, \label{bar-ep}
\end{eqnarray}
where $u(x,Q^2)$, $d(x,Q^2)$ etc.\ are parton distributions.

\begin{table}[b]
\caption[]{\label{table1}
The expected and observed numbers of events in each bin of the
$x-y$ plane from ZEUS. The expected number is listed above the observed
number.}
\arraycolsep=.5em
\bigskip
\[
\begin{array}{|r||c|c|c|c|c|c|c|c|c|}
\hline
x_{\rm min} & 0.05 & 0.15 & 0.25 & 0.35 & 0.45 & 0.55 & 0.65 & 0.75 &0.85  \\
x_{\rm max} & 0.15 & 0.25 & 0.35 & 0.45 & 0.55 & 0.65 & 0.75 &0.85  & 0.95\\
\hline
\hline
0.85 < y <0.95 & 8.8 & 1.2 & 0.32 & 0.10 & 0.028 & 0.01 & 0.0034 & & \\
               &  9  &  3  &      &      & 1     &      &        & & \\
\hline
0.75 < y <0.85 & 12  & 2.5 &  0.50& 0.15 & 0.050 & 0.011& 0.0039 & & \\
               & 16  & 4   &  1   &      &       &      &        & & \\
\hline
0.65 < y <0.75 & 13  & 3.7 & 0.86 & 0.26 & 0.082 & 0.022& 0.0054 &0.0020& \\
               & 10  & 3   &      &      &       &      &   1    &      & \\
\hline
0.55 < y <0.65 & 15  & 6.1 & 1.65 & 0.46 & 0.15  & 0.046& 0.0090 & 0.0024& \\
               & 12  & 3   &  3   &  1   &       &      &        &       & \\
\hline
0.45 < y <0.55 & 12  & 11  & 2.5  & 0.85 & 0.28  & 0.084& 0.0208 & 0.0032& \\
               & 6   & 13  &  1   &      &       &   1  &        &       &  \\
\hline
0.35 < y <0.45 & 4.6 & 18  & 5.5  & 1.75 & 0.52  & 0.16 & 0.0403 & 0.0093& \\
               & 3   & 17  &  6   &      &       &      &        &       & \\
\hline
0.25 < y <0.35 & & 18  & 11   & 3.74 & 1.19  & 0.34 & 0.1104 & 0.0175&0.0066 \\
               & & 23  & 6    &  7   & 1     & 2    &        &       &   \\
\hline
0.15 < y <0.25 & & 2.2 & 14   & 9.6  & 3.32  & 1.2  & 0.2784 &0.0717 &0.0077\\
               & & 1   & 15   & 10   & 3     &      & 1      & & \\
\hline
0.05 < y <0.15 & &     &      & 1.3  & 2.14  & 1.6  & 0.9052 &0.3022 &0.1216\\
               & &     &      & 1    & 3     & 2    & 1      & 1  & \\
\hline
\end{array}
\]
\end{table}

\begin{table}[h]
\caption[]{\label{table2}
The expected and observed numbers of events in each bin of the
$x-y$ plane from H1. The expected number is listed above the observed number.}
\arraycolsep=.5em
\medskip
\[
\begin{array}{|r||c|c|c|c|c|c|c|c|}
\hline
x_{\rm min} & 0.06236 & 0.2494 & 0.3395 & 0.3592 & 0.3794 & 0.4002 & 0.4216 &
0.4435 \\
x_{\rm max} & 0.2494 & 0.3395 & 0.3592 & 0.3794 & 0.4002 & 0.4216 & 0.4435 &
0.95 \\
\hline
\hline
0.5 < y < 0.9 & 73.59& 2.2  & 0.27 & 0.21 & 0.15 & 0.15 & 0.14 & 0.39 \\
              & 76   & 3    & 0    & 0    & 1    & 0    & 1    & 3 \\
\hline
0.4 < y < 0.5 & 56.28& 2.28 & 0.25 & 0.21 & 0.19 & 0.07 & 0.08 & 0.24 \\
              & 59   & 2    & 0    & 0    & 0    & 1    & 1    & 0   \\
\hline
0.3 < y < 0.4 & 65.13& 3.77 & 0.40 & 0.42 & 0.36 & 0.24 & 0.19 & 0.69 \\
              & 80   & 3    & 1    & 1    & 0    &   0  &  0   & 0   \\
\hline
0.2 < y < 0.3 & 70.9 & 9.93 & 1.05 & 0.91 & 0.75 & 0.61 & 1.45 & 1.4 \\
              & 64   & 12   & 2    & 0    & 1    & 1    & 0    & 1 \\
\hline
\end{array}
\]
\end{table}

\subsection{Atomic Parity Violation Experiment}

Parity violation in the SM is due to exchanges of weak gauge bosons, with
vector-axial-vector ($VA$) and axial-vector-vector ($AV$) terms contributing.
Atomic physics parity violation measures
\cite{langacker} electron-quark couplings that are different than those probed
by high energy experiments and thereby provides alternative constraints on new
physics.
In terms of the coefficients $C_{1q}$ and $C_{2q}$ of the NC Lagrangian
(\ref{lag}), the weak charge $Q_W$ of a heavy atom is given by~\cite{langacker}
\begin{equation}
Q_W = -2 \biggr[ C_{1u} (2Z+N) + C_{1d} (Z+2N) \biggr] \;,
\end{equation}
where $Z$ and $N$ are the number of protons and neutrons respectively in the
nucleus of the atom.
Recently a very precise measurement was made of the parity violation transition
between the 6S and 7S states of Cesium with the use of a spin-polarized atomic
beam\cite{apv}.
 For $^{133}_{\phantom055}$Cs, $ Q_{\rm w} \equiv
-376C_{1u}-422C_{1d}$.
The  weak charge $Q_W$ of the atom was determined to be
\begin{equation}
Q_W^{\rm exp} = -72.11\pm 0.27 \pm 0.89 \;, \label{weakcharge}
\end{equation}
where the first uncertainty is experimental and the second is theoretical.
The new measurement of (\ref{weakcharge}) represents substantial improvement
from the value given in the
1996 Particle Data Book\cite{pdg} and shows better agreement with the SM value.
Including radiative corrections the predicted SM value for Cs is\cite{chm97}
\begin{equation}
Q_W^{\rm SM}=-73.11 \pm 0.05 \,.
\end{equation}
We find the following constraint for the new physics contribution
\bq
\Delta Q_W(Cs) =
	-376 \Delta C_{1u} -422 \Delta C_{1d} = 1.00 \pm 0.93 \,.
\eq
Some chirality combinations of $LL,RR,LR,RL$ give zero
contributions to $\Delta C_{1q}$ and thus satisfy the experimental $Q_W$
constraint.  Such possibilities include (i) $LL=RR=LR=RL \;(VV)$,
(ii) $LL=RR=-LR=-RL\;(AA)$, (iii) $LL=-LR, RL=-RR$ (an SU(12) symmetry
 \cite{nelson}), (iv) $LR=RL, LL=RR=0$ (a minimal choice used in fitting the
HERA data \cite{ours}).

\subsection{Polarization Asymmetries in Electron-Nucleus Scattering Experiment}

In this subsection, we review three experiments on polarized
electron-nucleus scattering:
the SLAC $e$-D scattering experiment \cite{eD},
the Mainz $e$-Be scattering experiment \cite{mainz89},
and
the Bates $e$-C scattering experiment \cite{bates90}.

\subsubsection{SLAC $e$-D experiment }

The SLAC polarized $e$-D experiment \cite{eD} measured
the parity-violating
asymmetry
\begin{equation}
A = \frac{ d\sigma_R - d\sigma_L}{ d\sigma_R + d\sigma_L}\;,
\end{equation}
where $d\sigma_{R/L}$ are the differential cross sections for
$e^-_{R/L} {\rm D} \to e X$ scattering.  At low
energy $A$ is given in the valence approximation to the parton model
by\cite{pdg,langacker}
\begin{equation}
\frac{A}{Q^2} = \frac{3G_F}{5\sqrt{2}\pi \alpha_{\rm em}} \;
\Biggr[ \left (C_{1u} -\frac{C_{1d}}{2} \right ) +
        \left (C_{2u} -\frac{C_{2d}}{2} \right ) \;
     \frac{1-(1-y)^2}{1+(1-y)^2 } \Biggr ] \;,
\end{equation}
where $Q^2>0$ is the momentum transfer, $y$ is the fractional energy
transfer from the electron to hadrons, and $C_{1q}$, $C_{2q}$ are the
coefficients of the parity-violating Lagrangian in Eq.~(\ref{lag}).
After taking into account uncertainties in the sea-quark contribution,
the $R=\sigma_L/\sigma_R$ ratio and the higher-twist effects,
the following constraints are found \cite{hhkm94}
\begin{equation}
	2C_{1u}-C_{1d} = -0.94 \pm 0.26\,,
	\quad
	2C_{2u}-C_{2d} =  0.66 \pm 1.23\,,
        \quad
	\rho_{\rm corr} = -0.975\,,
\label{eq:ed_exp}
\end{equation}
where $\rho_{\rm corr}$ is the correlation.
The recent update of the SM contribution to the effective
parameters at $\langle Q^2 \rangle \simeq 1.5\ {\rm GeV}^2$
finds \cite{chm97}
\begin{mathletters}
\bea
(2C_{1u}-C_{1d})^{\rm SM} &=& -0.723 \pm 0.005, \\
(2C_{2u}-C_{2d})^{\rm SM} &=&  0.105 \pm 0.006.
\eea
\end{mathletters}
We hence use the following constraints for the new physics contributions
\begin{equation}
	\Delta( 2C_{1u}^e-C_{1d}^e) = -0.22 \pm 0.26 \,,
	\quad
	\Delta( 2C_{2u}^e-C_{2d}^e) = 0.77 \pm 1.23 \,,
	\quad
	\rho_{\rm corr} = -0.975\,.
\end{equation}
Note the strong negative correlation of the errors.

\subsubsection{Mainz $e$-Be experiment}

At Mainz, asymmetry of quasi-elastic polarized electron scattering
on a  $^9{\rm Be}$ target has been measured~\cite{mainz89}.
The experiment was performed at the mean momentum transfer
$\langle Q^2 \rangle \simeq 0.2025\ {\rm GeV}^2$.
The measured asymmetry parameter $A_{\rm Mainz}$ is given by
the following combination of the coefficients of the effective
Lagrangian (\ref{lag})
\beq
A_{\rm Mainz} = 2.73 C_{1u} - 0.65 C_{1d} + 2.19 C_{2u} - 2.03 C_{2d},
\eeq
and the experimental result is
\beq
A_{\rm Mainz} = -0.94 \pm 0.19. \label{A_mainz}
\eeq
Here the error is obtained by taking the quadratic sum of
the theoretical and experimental errors as quoted in Ref.~\cite{mainz89}.

The theoretical prediction in the SM has been evaluated in \cite{chm97} to be
\beq
A_{\rm Mainz}^{\rm SM} = -0.875 \pm 0.014.
\eeq
We hence find the following constraint on new physics contribution
\bea
\Delta A_{\rm Mainz} &=&
  2.73\Delta C_{1u}^e - 0.65\Delta C_{1d}^e
+ 2.19\Delta C_{2u}^e - 2.03\Delta C_{2d}^e \nonumber \\
&=&
-0.065 \pm 0.19.
\eea

\subsubsection{Bates $e$-C experiment}

The asymmetry of elastic polarized electrons scattering on a
$^{12}{\rm C}$ target has been measured at Bates~\cite{bates90}.
Their asymmetry parameter can be expressed in terms of the
effective Lagrangian coefficients as follows;
\begin{equation}
A_{\rm Bates} = \frac{3\sqrt{2} G_F Q^2}{4\pi\alpha(-Q^2) }
\Bigl( C_{1u} + C_{1d} \Bigr).
\end{equation}
The typical momentum transfer of the experiment is small,
$\langle Q^2 \rangle = 0.0225$ GeV$^2$, and contributions
from quarks other than the $u$ and $d$ quarks can be safely
neglected~\cite{beck87}. From the measurement\cite{bates90}
\beq
A_{\rm Bates} = (1.62 \pm 0.38) \times 10^{-6}, \label{A_bates}
\eeq
and the estimate \cite{burkhardt95}
$1/\alpha(-0.0225\ {\rm GeV}^2) = 135.87$,
we obtain the following constraint on the effective Lagrangian
coefficients,
\beq
C_{1u}+C_{1d} = 0.137 \pm 0.033.
\eeq
The theoretical prediction of the SM at
$\langle Q^2 \rangle = 0.0225\ {\rm GeV}^2$ is \cite{chm97}
\beq
(C_{1u}+C_{1d})^{\rm SM} = 0.1522 \pm 0.0004.
\eeq
Therefore the constraint on new physics is
\beq
\Delta (C_{1u}^e + C_{1d}^e) = -0.0152 \pm 0.033.
\eeq

\subsection{CDF Drell-Yan Cross Sections}

The CDF experiment at the Tevatron has reported preliminary data \cite{cdf-dy}
on Drell-Yan lepton-pair production.  The data that we use in our analysis
are extracted from a CDF figure; the values are given in Table \ref{table4}.
The data are
differential cross sections versus $M$, integrated over $|y|<1$ and
divided by two to average over rapidity,
where  $M$ and $y$ are, respectively, the invariant mass and rapidity of the
lepton pair.
The double differential cross section versus $M$ and $y$
of the lepton pair is given by
\begin{equation}
\frac{d^2\sigma}{dM dy} =K\, \frac{M^3}{72\pi s} \sum_q q(x_1) \bar q(x_2)
\biggr[
|M_{LL}^{eq}(\hat s)|^2 + |M_{RR}^{eq}(\hat s)|^2 + |M_{LR}^{eq}(\hat s)|^2
+ |M_{RL}^{eq}(\hat s)|^2 \biggr] \;,
\end{equation}
where $q(x)$ and $\bar q(x)$ are  parton distributions evaluated at
$Q^2=M^2$, $x_{1,2} = M e^{\pm y}/\sqrt{s}$,  $\sqrt{s}=1800$ GeV, and
$\hat s = M^2$.
The reduced amplitudes $M_{\alpha\beta}^{eq}(\hat s)$ are given by
Eq.~(\ref{reduced}) with $\hat t$ replaced by $\hat s$.
The QCD $K$-factor of Drell-Yan production is given by\cite{coll-phys}
\begin{equation}
K = 1 + \frac{\alpha_s(M^2)}{2\pi} \,\frac{4}{3} \left(
1+ \frac{4}{3}  \pi^2 \right ) \;.
\end{equation}
With this $K$ factor, the overall cross section normalization
agrees with the CDF data in the vicinity of the $Z$-peak.

\begin{table}[h]
\caption[]{\label{table4}
Preliminary CDF Drell-Yan Data extracted from their figures.  Both
$e^+ e^-$ and $\mu^+\mu^-$ samples are shown. The data is the $y$-averaged
integrated cross section
$ \frac{1}{2} \int_{-1}^{1} dy \frac{d^2 \sigma}{dM dy}$. }
\medskip
\arraycolsep=0.5em
\[
\begin{array}{|c|c||c|c|}
\hline
M_{ee} ({\rm GeV}) & \frac{1}{2} \int_{-1}^1 dy \frac{d^2 \sigma}{dM dy}
({\rm pb}/{\rm GeV} )   &
M_{\mu\mu} ({\rm GeV}) & \frac{1}{2} \int_{-1}^1 dy \frac{d^2 \sigma}{dM dy}
({\rm pb}/{\rm GeV} ) \\
\hline
\hline
54.63 & 0.142 \stackrel{\scriptstyle +0.024}{\scriptstyle -0.023} &
54.63 & 0.118 \pm {0.019} \\
64.90 & 0.113 \stackrel{ +0.017}{\scriptstyle -0.016} &
64.90 & 0.098 \stackrel{+0.015}{\scriptstyle -0.014} \\
73.82 & 0.130 \stackrel{ +0.016}{\scriptstyle -0.017} &
73.82 & 0.113 \stackrel{+0.017}{\scriptstyle -0.016} \\
81.86 & 0.358 \stackrel{ +0.029}{\scriptstyle -0.035} &
81.86 & 0.329 \stackrel{+0.029}{\scriptstyle -0.030} \\
87.66 & 2.537 \stackrel{ +0.168}{\scriptstyle -0.188} &
87.66 & 2.162 \stackrel{+0.158}{\scriptstyle -0.173} \\
91.68 & 10.779 \stackrel{ +0.861}{\scriptstyle -0.668}&
91.68 & 11.345 \stackrel{+0.906}{\scriptstyle -0.839} \\
97.92 & 0.462 \stackrel{+0.037}{\scriptstyle -0.045} &
97.92 & 0.502 \stackrel{+0.044}{\scriptstyle -0.040} \\
105.96 & 0.122 \stackrel{+0.022}{\scriptstyle -0.020} &
107.74 & 0.148 \stackrel{+0.020}{\scriptstyle -0.021} \\
114.88 & 0.0427 \stackrel {+0.0104}{\scriptstyle -0.0105} &
116.67 & 0.0635 \stackrel{+0.0125}{\scriptstyle -0.0131} \\
134.97 & 0.0209 \pm{0.0041} &
138.98 & 0.0214 \stackrel{+0.0045}{\scriptstyle -0.0042} \\
174.69 & 0.00352 \pm{0.00127} &
178.71 & 0.00426 \stackrel{+0.0015}{\scriptstyle -0.0014} \\
225.11 & 0.00234 \pm{0.00105} &
228.69 & 0.00195 \stackrel{+0.00099}{\scriptstyle -0.00097} \\
275.11 & (9.41 \pm{6.70})\times 10^{-4} &
279.13 & (9.91 \pm{7.05})\times 10^{-4} \\
350.09 & \left(2.22 \stackrel{+2.21}{\scriptstyle -2.22} \right)\times 10^{-4}&
354.11 &\left(2.22 \stackrel{+2.26}{\scriptstyle -2.22} \right)\times 10^{-4}\\
450.06 & (2.27 \pm 2.27 )\times 10^{-4} &
454.08 & \left(0.0 \stackrel{+2.03}{\scriptstyle -0.0}\right) \times 10^{-4} \\
550.04 & \left(0.0 \stackrel{+2.27}{\scriptstyle -0.0}\right) \times 10^{-4} &
554.06 & \left(0.0 \stackrel{+1.74}{\scriptstyle -0.0} \right)\times 10^{-4} \\
\hline
\end{array}
\]
\end{table}

\subsection{$e^+e^-$ Experiments}

The $eeqq$ contact interactions probed at HERA and in low energy experiments
involve only light quarks. In $e^+e^-\to{}$hadrons both light and heavy quarks
contribute and separation of the light quark
contributions requires bottom and charm quark tagging. To investigate the
influence of contact terms on $e^+e^-\to\rm hadrons$ we assume here that the
$eeqq$ contact interaction is flavor independent.
The relevant measurements
are the total hadronic cross sections
$\sigma_{\rm had}$ and the left-right asymmetry $A_{LR}$ at LEP1 and SLD and
$\sigma_{\rm had}$  at LEP1.5 and
LEP2.  Our analysis is based on the
data in Refs.~\cite{lep,opal,L3,aleph,langacker2}.

At leading order in the electroweak interactions the total hadronic cross
section  summed over all flavors $q=u,d,s,c,b$  is
\begin{equation}
\sigma_{\rm had} = K\, \sum_q \frac{s}{16\pi} \biggr[
|M_{LL}^{eq}(s)|^2 + |M_{RR}^{eq}(s)|^2 + |M_{LR}^{eq}(s)|^2 +
|M_{RL}^{eq}(s)|^2 \biggr] \,,
\label{sig_had}
\end{equation}
where the QCD $K$ factor is given by $K = 1 + \alpha_s/\pi +
1.409(\alpha_s/\pi)^2
-12.77(\alpha_s/\pi)^3$\cite{gorishny91}.
The left-right asymmetry $A_{LR}$ is given by
\begin{equation}
A_{LR} = \frac{\sigma_L - \sigma_R}{\sigma_L + \sigma_R}
       = \frac{\sum_q \left(|M_{LL}^{eq}(s)|^2 + |M_{LR}^{eq}(s)|^2
                     -|M_{RL}^{eq}(s)|^2 - |M_{RR}^{eq}(s)|^2 \right)}
               {\sum_q \left( |M_{LL}^{eq}(s)|^2 + |M_{LR}^{eq}(s)|^2
                     +|M_{RL}^{eq}(s)|^2 + |M_{RR}^{eq}(s)|^2 \right)} \,.
\label{assym}
\end{equation}
At $\sqrt{s}=M_Z$,  the SM amplitudes are imaginary whereas contact term
contributions are real. Because of the absence of interference with the SM
amplitudes, the $Z$-pole data have little sensitivity to the contact terms
despite their high accuracy.
To take into account next-to-leading order (NLO) electroweak radiative
corrections we calculate
\begin{equation}
\sigma_{\rm had}^{\rm theory} = \sigma_{\rm LO}^{\rm theory} \times \left(
\sigma^{\rm SM}_{\rm NLO}\over \sigma_{\rm LO}^{\rm SM} \right) \,,
\end{equation}
where $\sigma_{\rm LO}^{\rm theory}$ is given by (\ref{sig_had}) with contact
interaction contributions included in the matrix elements. Table~\ref{table5}
gives the measured $\sigma_{\rm had}$ and the calculated values of $\sigma_{\rm
NLO}^{\rm SM}$.
The radiative corrections to the LR asymmetry cancel in the ratio (\ref{assym})
so here we
compare the observed $A_{LR}$ with the leading order SM calculation.
The experimental and the SM predicted values for $A_{LR}$ are \cite{lep}
\begin{equation}
\label{alr}
A_{LR}^{\rm exp} = 0.1542 \pm 0.0037\;, \qquad A_{LR}^{\rm SM}=0.145 \pm
0.001 \pm 0.001\;.
\end{equation}
Other LEP1.5 and LEP2 measurements are summarized in Table \ref{table5}.
The new physics contributions are given by Eq.~(\ref{newphys}).

\begin{table}[h]
\caption[]{\label{table5}
Table showing the LEP1 and SLD data, and LEP1.5 and LEP2 data
that are relevant to our analysis \cite{lep,opal,L3,aleph,langacker2}.}
\medskip
\arraycolsep=0.75em
\[
\begin{array}{|l||c|c|}
\hline
\multicolumn{1}{|c||}{E_{CM}} &  \sigma_{\rm had} &\sigma_{\rm had}^{\rm SM} \\
\hline
\hline
{\rm LEP1,SLD:}
\sqrt{s}=M_Z               & 41.508\pm0.056 \;{\rm nb} & 41.465\; {\rm nb} \\
\hline
{\rm OPAL: }\sqrt{s}=130.26\;{\rm GeV} & 66\pm 5\pm 3 \;{\rm pb} & 78\;{\rm
pb}\\
{\rm OPAL: }\sqrt{s}=136.23\;{\rm GeV} & 60\pm 5 \pm 2 \;{\rm pb} &63\;{\rm
pb}\\
{\rm OPAL: }\sqrt{s}=140\;{\rm GeV} &    50\pm 36 \pm 2\;{\rm pb} &56\;{\rm
pb}\\
{\rm L3: }\sqrt{s}=130.3\;{\rm GeV} & 81.8\pm6.4 \;{\rm pb} & 78\; {\rm pb}\\
{\rm L3: }\sqrt{s}=136.3\;{\rm GeV} & 70.5\pm6.2 \;{\rm pb} & 64\; {\rm pb}\\
{\rm L3: }\sqrt{s}=140.2\;{\rm GeV} & 67\pm47 \;{\rm pb} & 56\; {\rm pb}\\
{\rm ALEPH: }\sqrt{s}=130\;{\rm GeV} & 74.2\pm5.2\pm3.3 \;{\rm pb} & 76.9\;{\rm
pb}\\
{\rm ALEPH: }\sqrt{s}=136\;{\rm GeV} & 57.4\pm4.5\pm1.8 \;{\rm pb} & 62.5\;
{\rm pb}\\
\hline
{\rm OPAL: }\sqrt{s}=161.3\;{\rm GeV} & 35.3\pm 2.0 \pm 0.7\;{\rm pb} & 33.2\;
{\rm pb}\\
{\rm L3: }\sqrt{s}=161.3\;{\rm GeV} & 37.3\pm 2.2 \;{\rm pb} & 34.9\;{\rm pb}\\
{\rm L3: }\sqrt{s}=170.3\;{\rm GeV} & 39.5\pm 7.5 \;{\rm pb} & 29.8\;{\rm pb}\\
{\rm L3: }\sqrt{s}=172.3\;{\rm GeV} & 28.2\pm 2.2 \;{\rm pb} & 28.9\;{\rm pb}\\
\hline
\end{array}
\]
\end{table}

\subsection{Neutrino-Nucleon DIS Experiments}

Deep inelastic scattering (DIS) experiments with neutrino and anti-neutrino
beams have provided important tests for the SM since the early
80's \cite{haidt,global}.
If SU(2)$_L$ invariance is assumed, then $\nu N$ DIS data also constrain
contact interactions.
The CCFR collaboration obtained a model-independent constraint on the effective
$\nu\nu qq$ couplings\cite{ccfr}:
\begin{equation}
\kappa = 0.5820 \pm 0.0041 = 1.7897 g_L^2 + 1.1479 g_R^2 - 0.0916 \delta_L^2
 - 0.0782 \delta_R^2 \;, \label{ccfr_kappa}
\end{equation}
where $g_{L,R}^2 = \left(g_{L,R}^u\right)^2+\left(g_{L,R}^d\right)^2$,
$\delta_{L,R}^2 =
(g_{L,R}^u)^2 - (g_{L,R}^d)^2$ and $g_L^q,g_R^q$ are
the coefficients of the effective Lagrangian (\ref{effLnuq}).  The Standard
Model value is $\kappa=\kappa_{\rm
SM} = 0.5817 \pm 0.0013$ \cite{ccfr}.
This model-independent constraint (\ref{ccfr_kappa}) can be used to constrain
physics beyond the SM.
Since this CCFR result is so far the most accurate measurement for $\nu N$
scattering, we adopt their result in our analysis.
We can write $g_L^u = (g_L^u)_{\rm SM} + \Delta g_L^u$, etc., then
$\kappa$ becomes
\begin{eqnarray}
\kappa &=& \kappa_{\rm SM} + 1.7897 \left(
 2 (g_L^u)_{\rm SM} \Delta g_L^u +(\Delta g_L^u)^2 +
 2 (g_L^d)_{\rm SM} \Delta g_L^d +(\Delta g_L^d)^2  \right )  \nonumber \\
&+& 1.1479 \left(
 2 (g_R^u)_{\rm SM} \Delta g_R^u +(\Delta g_R^u)^2 +
 2 (g_R^d)_{\rm SM} \Delta g_R^d +(\Delta g_R^d)^2  \right )  \nonumber \\
&-& 0.0916 \left(
 2 (g_L^u)_{\rm SM} \Delta g_L^u +(\Delta g_L^u)^2 -
 2 (g_L^d)_{\rm SM} \Delta g_L^d -(\Delta g_L^d)^2  \right )  \nonumber \\
&-& 0.0782 \left(
 2 (g_R^u)_{\rm SM} \Delta g_R^u +(\Delta g_R^u)^2 -
 2 (g_R^d)_{\rm SM} \Delta g_R^d -(\Delta g_R^d)^2  \right )  \;, \label{kapp}
\end{eqnarray}
where  the $\Delta g_{L,R}^{u,d}$ are the contact term
contributions of Eq.~(\ref{shift}).

\subsection{Statistical Analysis}

The method of maximum likelihood is the most general method of parameter
estimation.
For a set of independently measured quantities $x_i$ that come from
a probability density function $f(x;\alpha)$, where $\alpha$ is a set of
unknown parameters,  the method of maximum likelihood consists of finding
the set of $\hat \alpha$, which maximizes the joint probability density
for all data, given by the likelihood function
${\cal L} = \prod_i f(x_i;\alpha)$.   Very often it is easier to
maximize the logarithm of this likelihood function by solving the
equation $\partial \log {\cal L}/\partial \alpha=0$.
In the case of ZEUS and H1 data, since the number of observed events in each
bin is small, it is appropriate to describe the
probabilities of the observed events using Poisson statistics:
\begin{equation}
f_i = \frac{ {(n_i^{\rm th})}^{n_i^{\rm obs} } e^{-n_i^{\rm th}} }
{n_i^{\rm obs} !} \;,
\end{equation}
where $n_i^{\rm th}$ is the expected number of events from the theory and
$n_i^{\rm obs}$ is the number of observed events in the $i$th bin.
With Poisson statistics the method of maximum log-likelihood is
equivalent to minimizing the following $\chi^2$ \cite{pdg}
\begin{equation}
\label{chi2}
\chi^2 = \sum_i \biggr [
2(n_i^{\rm th} - n_i^{\rm obs}) + 2 n_i^{\rm obs} \log \left(
\frac{n_i^{\rm obs}}{n_i^{\rm th}} \right) \biggr ] \;.
\end{equation}
In the bins where $n_i^{\rm obs}=0$ the second term is zero.  In our fitting
procedures,  we shall vary unknown $n$ parameters of the contact interaction
and calculate the
expected number of events, $n_i^{th}$, for each bin $i$; then we obtain
the quantity $\chi^2$ for all relevant ranges of the unknown parameters.
The set of parameters that gives the minimum $\chi^2$ value is the best
estimate of the parameters.
Relative merits of different new physics interactions can also be
compared by their $\chi^2$ values.

In experiments with good statistics, we use the usual
approach to calculate
\begin{equation}
\chi^2 = \sum_{i=\mbox{\scriptsize bins or data points}} \left(
\frac{x_i^{\rm exp} - x_i^{th}}{\sigma_{x_i}} \right )^2 \;,
\end{equation}
where $x_i^{exp}$, $x_i^{th}$, and $\sigma_{x_i}$ are the experimental
measurements, theoretical predictions, and the errors, respectively.
The method of maximum log likelihood is equivalent to the usual
$\chi^2$ method when the distribution follows Gaussian statistics.
The $\chi^2$'s from all data sets are added together to form
the total $\chi^2$ and
best estimate $\hat \alpha$ for the set of parameters is obtained
by minimizing the total $\chi^2$, i.e., $\chi^2(\hat\alpha)=\chi^2_{\rm min}$.
The $n$-sigma standard deviation of the best estimate $\hat \alpha$ is
calculated by finding $\hat \alpha'$, which satisfies \cite{pdg}
\begin{equation}
\chi^2(\hat \alpha') = \chi^2_{\rm min} + n^2 \;.
\end{equation}
For a 1(2)-parameter fit a 1-$\sigma$ standard deviation corresponds to
a confidence level of 68(39)\%.

\section{\lowercase{$eeqq$} Contact Interactions}

In this section we present the results of our global fit to the $eeqq$ contact
interactions. The HERA data, Tables~\ref{table1} and \ref{table2}, mainly
constrain the $eeuu$ and $eedd$ contact interactions, as do the cesium atom
weak charge (\ref{weakcharge}) and the three asymmetries (\ref{eq:ed_exp}),
(\ref{A_mainz}), (\ref{A_bates}) of polarized electron scattering on nuclei
targets. The Drell-Yan (DY) lepton-pair production data, Table~\ref{table4},
are sensitive to $eeqq$ and $\mu\mu qq$ contact interactions mainly of the
first generation quarks ($u$ and $d$). The DY data show no indication of
$e$-$\mu$ universality violation, hence we assume the $e$-$\mu$ universality of
the contact interactions when we include this data in our fit. The LEP/SLC
measurements of $e^+e^-\to{}$hadrons, Table~\ref{table5} and Eq.~(\ref{alr})
constrain $eeqq$
contact interactions summed over five flavors of quarks, $u,d,s,c$ and $b$.
Since $b$ and $c$ flavor-tagging measurements have only limited accuracy,
we need to assume quark-flavor universality relations when we include the
LEP/SLC data in our global fit,
\begin{equation}
\eta_{\alpha\beta}^{eu} = \eta_{\alpha\beta}^{ec}\,,\quad
\eta_{\alpha\beta}^{ed} = \eta_{\alpha\beta}^{es} = \eta_{\alpha\beta}^{eb}\,,
\end{equation}
where $\alpha\beta= LL, LR, RL, RR$. Finally, the neutrino-nucleon scattering
data (\ref{ccfr_kappa}) constrain the neutral current $\nu_\mu\nu_\mu uu$ and
$\nu_\mu\nu_\mu dd$ contact interactions. To include neutrino data in the
global analysis we need to assume the
electroweak gauge symmetry relations of Eq.~(\ref{su2relnu}). Our global fits
for the $eeqq$ contact interactions are thereby organized in the increasing
order of model dependence.

In Table~\ref{table6} we summarize the experimental constraints obtained on
the $eeqq$ contact interactions $\eta_{\alpha\beta}^{eq}$ in the various
``models":
(a)~the SM result,
(b)~the choice $\eta_{LR}^{eu} = \eta_{RL}^{eu} = 1.4\rm~GeV^{-2}$ used in
our previous qualitative demonstration of the effects of the contact
interactions\cite{ours},
(c)~the fit when all 8 $eeqq$ contact terms are allowed to vary freely,
(d)~a minimal choice  that would explain the HERA and other data,
(e)~a vector-vector model,
(f)~an axial vector - axial vector model, and
(g)~the SU(12) model\cite{nelson}.
The fits were performed using the program MINUIT\cite{minuit}.

The breakdown of the minimum $\chi^2$ for each model is
summarized in Table \ref{table6b}.
The SM global-fit is excellent,
giving $\chi^2/\mbox{d.o.f.} = 176.4/164$. The SM gives the best $\chi^2$ of
the CDF-DY data. Only for the H1 data does the SM give a poor representation.
Our previous choice (b) fits the H1 data but does not fit ZEUS nor CDF-DY
data well.
For case (c) in which all eight
parameters are allowed to vary, the minimization maintains a balance among
the $\chi^2$ of the data sets.  From Tables \ref{table6} and \ref{table6b}
we see that
the best estimate for  the  $LR$ and $RL$ components are similar and the
minimum $\chi^2$ for (c) and (d) also is comparable.
This confirms that the HERA anomaly is better
explained by  $LR$ and $RL$ components, which have less influence on
other experiments.
In the cases (e), (f), and (g) the $\chi^2$ for the HERA data does not
improve over (c) and (d), which indicates that
the $VV$, $AA$, and the SU(12) chirality combinations cannot better explain
the HERA anomaly without upsetting other experiments.  Other
$\eta_{\alpha\beta}^{eq}$ solutions could conceivably give a smaller $\chi^2$
for HERA but
only at the expense of larger $\chi^2$'s for the other experiments.
The above fits in Table~\ref{table6} and \ref{table6b} do not include the $\nu
N$ data because most of the ``models" do not maintain the SU(2) invariance
relations (\ref{su2releR}) and (\ref{su2relnu}).

In all the fits, we notice that $\eta_{LR}^{ed}$ and $\eta_{RL}^{ed}$
are in general larger than the others, even larger than
$\eta_{LR}^{eu}$ and $\eta_{RL}^{eu}$.
This is because (i) the excess events at HERA is best explained by
by the $LR$ and $RL$ chiralities (see Eq. (\ref{bar-ep})), (ii) in Drell-Yan
production at the Tevatron $u \bar u$ annihilation dominates over
$d \bar d$, and the Drell-Yan cross sections agree with the
SM prediction.  Therefore, the minimization process
increases $\eta_{LR}^{ed}$ and $\eta_{RL}^{ed}$ to fit the HERA data
without spoiling the SM fit to the Drell-Yan cross section.

Table \ref{table6c} shows how the best estimates
for $\eta_{\alpha\beta}^{eq}$ parameters change when data sets are added
successively in the minimization.  The most dramatic changes
occur when the data for Drell-Yan (DY) lepton-pair production at the Tevatron
are included.
Note that when we include the $\nu$-N data in the last column of Table
\ref{table6c} the SU(2)$_L$ invariance relations
of Eqs.~(\ref{su2releR}) and (\ref{su2relnu}) are assumed and then we have 7
free parameters instead of 8.
Table~\ref{table6c} shows that  the HERA and DY data are not fully compatible.
There is an abrupt change of the best-fit $\eta_{\alpha\beta}^{eq}$ values and
a sudden increase of the $\chi^2$ from the HERA data between the third
(HERA+APV+$e$N) and fourth (HERA+APV+$e$N+DY) columns. The difference
$\chi^2_{\rm SM} - \chi^2_{\rm min}$ is as large as 12.8 at the third column
but it reduces to 8.2 in the fourth column after including the CDF-DY data,
which is no significant improvement over the SM since there are 8 $eeqq$
contact terms allowed to vary freely.

\begin{figure}
\centering
\hspace{0in}\epsfxsize=6in\epsffile{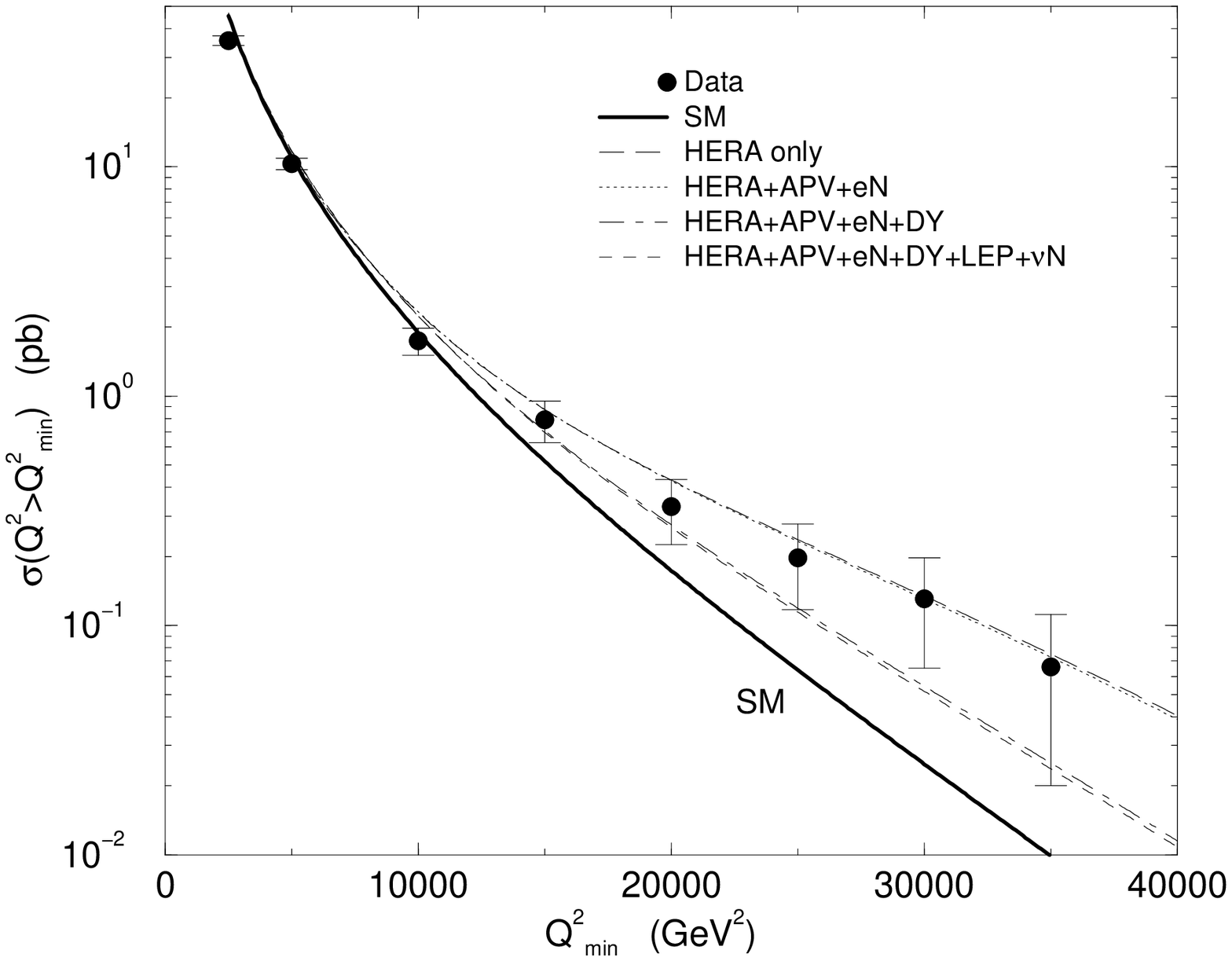}

\caption[]{\label{fig:ics}
Integrated cross sections for $e^+p\to e^+X$ versus a minimum $Q^2$ for
the SM (solid curve) and for four choices of contact interactions:
best-fit for the H1 and ZEUS data (long-dashed curved), best-fit including the
low energy electron-quark coupling measurements (dotted curve), best-fit
further including the Tevatron lepton-pair production data (dot-dashed curve),
and the best-fit including also the LEP data and the neutrino-scattering data
(dashed curve); see Table \ref{table6c}. The data points correspond to 
the number of events observed by H1\cite{H1} and ZEUS\cite{zeus}, and have 
been converted to cross sections by us, assuming constant detection 
efficiencies of 80\% and 81.5\%, respectively. 
Our fits are performed directly on the published
data\cite{H1,zeus} as summarized in Tables~\ref{table1} and \ref{table2}.}
\end{figure}

\begin{figure}
\centering
\hspace{0in}\epsfxsize=6in\epsffile{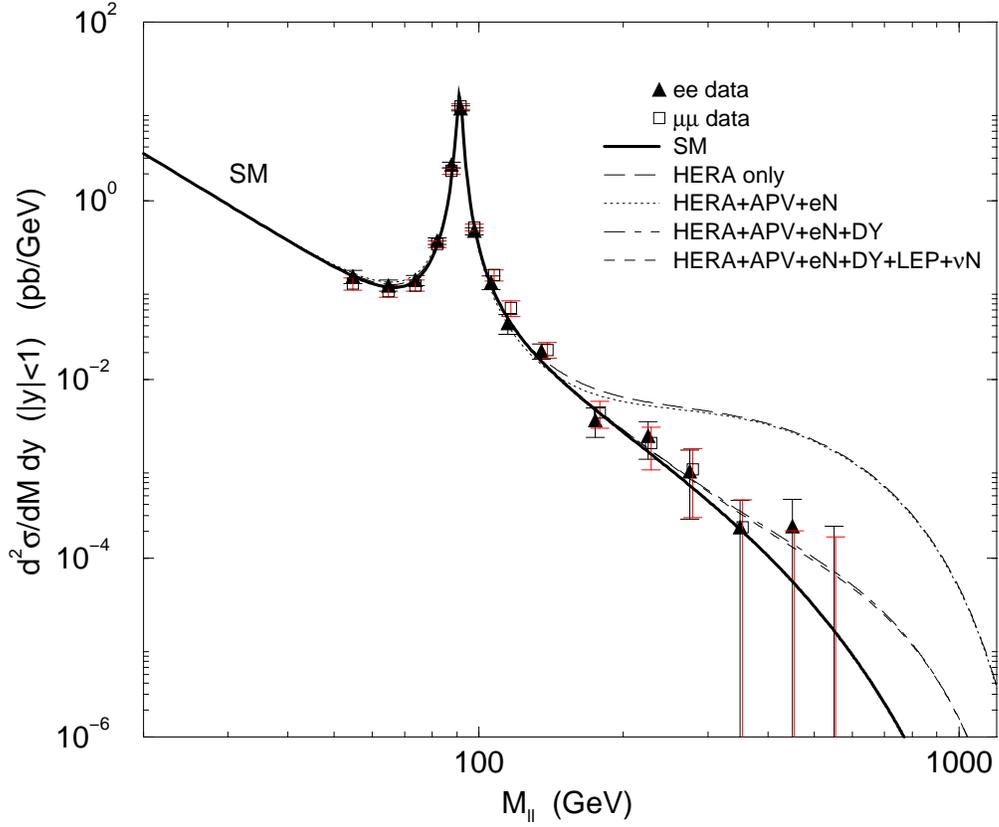}

\caption[]{\label{fig:DY} The Drell-Yan cross section $\bar pp\to
(\mu^+\mu^-\mbox{ or } e^+e^-)X$ at the Tevatron ($\sqrt s = 1.8$~TeV) for the
SM (solid curve) and the four choices of contact interactions as in
Fig.~\ref{fig:ics}. Preliminary CDF data\cite{lep} are shown separately for
$e^+e^-$ (solid triangle) and $\mu^+\mu^-$ (open square) pair; see
Table~\ref{table4}. The cross section is averaged over $y$ as ${1\over2}
\int_{-1}^{1} dy {d^2\sigma\over dMdy}$.}
\end{figure}

Figure~\ref{fig:ics} shows the fit
for the HERA data and Fig.~\ref{fig:DY} shows that for the CDF-DY data
from the models as given in Table~\ref{table4}:
(i)~fit to HERA only (long-dashed curves),
(ii)~fit to HERA+APV+$e$N data (dotted curves),
(iii)~fit including the DY data assuming $e$-$\mu$ universality
(dash-dotted curves),
(iv)~fit including the LEP and $\nu$N data, assuming the quark flavor
universality and SU(2)$_L$ invariance relations (dashed curves).
The incompatibility of the HERA high-$Q^2$ data and the CDF high mass DY
data can be seen in these figures.  Even allowing for general $eeqq$ contact
interactions, it is not possible to find a good fit to both the HERA and
CDF-DY data.

Finally, we present the result of the global fit to all low and high energy
lepton-quark experiments by assuming the flavor universality and the
SU(2)$_L$ invariance of the contact interactions. We then have the seven
independent lepton-quark contact interactions that appear in the last column of
Table~\ref{table6c}. The errors given for each $\eta_{\alpha\beta}^{eq}$ are
correlated. The correlation matrix as obtained from MINUIT\cite{minuit} is as
follows:

\begin{equation}
\left( { \scriptstyle \begin{array}{rrrrrrr}
  1.00&  -0.65&   0.78&  -0.52&   0.91&   0.14&  -0.13 \\
 -0.65&   1.00&  -0.76&   0.69&  -0.38&  -0.31&   0.47 \\
  0.78&  -0.76&   1.00&  -0.71&   0.58&   0.36&  -0.49 \\
 -0.52&   0.69&  -0.71&   1.00&  -0.43&   0.22&   0.75\\
  0.91&  -0.38&   0.58&  -0.43&   1.00&  -0.18&  -0.10 \\
  0.14&  -0.31&   0.36&   0.22&  -0.18&   1.00&   0.35 \\
 -0.13&   0.47&  -0.49&   0.75&  -0.10&   0.35&   1.00
            \end{array}
} \right ) \;,
\end{equation}
where the row and column indices are labeled in the order: $\eta_{LL}^{eu}$,
$\eta_{LR}^{eu}$, $\eta_{RL}^{eu}$, $\eta_{RR}^{eu}$, $\eta_{LL}^{ed}$,
$\eta_{LR}^{ed}$,  $\eta_{RR}^{ed}$.
The covariance matrix $C$ can be formed by $C_{ij} = \sigma_i \sigma_j
\rho_{ij}$, where $\sigma_i$ is the error of the $i$-th parameter.
The eigenvectors of the covariance matrix are constrained as:
\begin{mathletters} \label{eigvec}
\begin{eqnarray}
.386\eta_{LL}^{eu} + .319\eta_{LR}^{eu} - .620\eta_{RL}^{eu} -
.293\eta_{RR}^{eu} + .261\eta_{LL}^{ed} + .321\eta_{LR}^{ed} -
.329\eta_{RR}^{ed} &=& 0.034\pm 0.031 \,, \label{eigvec1}\\
.737\eta_{LL}^{eu} + .131\eta_{LR}^{eu} + .126\eta_{RL}^{eu} +
.080\eta_{RR}^{eu} - .630\eta_{LL}^{ed} - .133\eta_{LR}^{ed} +
.055\eta_{RR}^{ed}  &=& 0.014\pm 0.063\,, \label{eigvec2}\\
.216\eta_{LL}^{eu} - .545\eta_{LR}^{eu} - .232\eta_{RL}^{eu} +
.682\eta_{RR}^{eu} + .153\eta_{LL}^{ed} - .014\eta_{LR}^{ed} -
.338\eta_{RR}^{ed}  &=& -0.92 \pm 0.29\,, \label{eigvec3}\\
.035\eta_{LL}^{eu} - .678\eta_{LR}^{eu} - .394\eta_{RL}^{eu} -
.473\eta_{RR}^{eu} - .199\eta_{LL}^{ed} - .047\eta_{LR}^{ed} +
.343\eta_{RR}^{ed}  &=& -0.45\pm 0.37\,, \\
.429\eta_{LL}^{eu} + .041\eta_{LR}^{eu} + .096\eta_{RL}^{eu} +
.070\eta_{RR}^{eu} + .638\eta_{LL}^{ed} - .230\eta_{LR}^{ed} +
.583\eta_{RR}^{ed}  &=&  \phantom+0.33\pm0.82\,,\\
.063\eta_{LL}^{eu} - .098\eta_{LR}^{eu} + .187\eta_{RL}^{eu} +
.158\eta_{RR}^{eu} - .053\eta_{LL}^{ed} + .904\eta_{LR}^{ed} +
.326\eta_{RR}^{ed} &=&  \phantom+1.04\pm 1.46\,, \\
.268\eta_{LL}^{eu} - .336\eta_{LR}^{eu} + .590\eta_{RL}^{eu} -
.434\eta_{RR}^{eu} + .249\eta_{LL}^{ed} + .079\eta_{LR}^{ed} -
.459\eta_{RR}^{ed}  &=& -0.36 \pm 1.50\,.
\end{eqnarray}
\end{mathletters}
The strongest constraint (\ref{eigvec1}) is essentially due to the atomic
parity violation experiment and hence it agrees roughly with the result of the
corresponding fit to the low energy electroweak data only\cite{chm97}.
The rest of the constraints show significant improvements over those
obtained from low energy experiments, even though the analysis of
Ref.~\cite{chm97} made a more restrictive assumption that
$\eta_{LL}^{eu} = \eta_{LL}^{ed}$. The combined effect of low-energy and
high-energy data is that irrespective of the flavor and chirality
combination no $eeqq$ contact interactions can be significantly larger
than about 1~TeV$^{-2}$.

Throughout, we have parameterized the $eeqq$ contact interactions via the
$\eta^{eq}_{\alpha\beta}$ which, to a reasonable approximation, are normally
distributed. Constraints on contact terms are often quoted in terms of 95\%
CL bounds on the contact scale $\Lambda_{\alpha\beta\epsilon}^{eq}$
which is related to the $\eta$'s via
\begin{equation}
\eta_{\alpha\beta}^{eq} = \frac{4\pi \epsilon}{\left(
 \Lambda_{\alpha\beta\epsilon}^{eq} \right)^2 }
\end{equation}
where $\epsilon=\pm1, \alpha,\beta=L,R$.
In order to allow an easy comparison we also derive 95\% CL bounds on the
$\Lambda_{\alpha\beta}^{eq}$. The procedures for obtaining these limits
is as follows:
\begin{enumerate}

\item
Using MINUIT we obtain the best estimate of a parameter
$\eta_{\alpha\beta}^{eq}$ (others set to zero)
by minimizing the $\chi^2$ for all the available data.
When obtaining $\eta_{\alpha\beta}^{eq}= \mu \pm \sigma$, we
assume that the probability density function $P(\eta)$ follows
a normal distribution with mean $\mu$ and variance $\sigma^2$:
\begin{equation}
P(\eta) = \frac{1}{\sigma \sqrt{2\pi} }\; e^{- \frac{1}{2}
\left( \frac{\eta-\mu}{\sigma} \right )^2 } \;\;.
\end{equation}
Possible non-Gaussian behaviour is taken into account by replacing
$(\frac{\eta -\mu}{\sigma})^2$ by $\chi^2(\eta)-\chi^2_{\rm min}$ in the
argument of the exponential.

\item
We determine ``$+$'' and ``$-$''-sided limits of
$\Lambda_{\alpha\beta\epsilon}^{eq}$.
The ``$+$''-sided limit means that the physically allowed region of
$\eta_{\alpha\beta}^{eq}$ is $\eta_{\alpha\beta}^{eq} \ge 0$.
To obtain a 95\% CL limit on $\eta_{\alpha\beta}^{eq}$ we need to find
the value, $\eta_+$, above which 5\% of the total probability for positive
$\eta$ is located,
\begin{equation}
0.05 = \frac{\int_{\eta_+}^{\infty} P(\eta) d\eta }
              {\int_{0}^{\infty} P(\eta) d\eta } \; .
\end{equation}
This $\eta_+$ is related to $\Lambda_{+}$ by
$\eta_+=4\pi/(\Lambda_{+})^2$.

\item
The ``$-$''-sided limit means that the physically allowed region of
$\eta_{\alpha\beta}^{eq}$ is $\eta_{\alpha\beta}^{eq} < 0$.
To obtain a 95\% CL limit on $\eta_{\alpha\beta}^{eq}$ we need to find
the value, $\eta_-$, below which 5\% of the total probability for negative
$\eta$ is located,
\begin{equation}
0.05 = \frac{\int_{-\infty}^{\eta_-} P(\eta) d \eta }
              {\int_{-\infty}^{0} P(\eta) d \eta }\; .
\end{equation}
This $\eta_-$ is related to $\Lambda_{-}$ by
$\eta_-= - 4\pi/(\Lambda_{-})^2$.

\end{enumerate}
The limits are tabulated in Table~\ref{table-new1}
for individual couplings of a fixed
lepton and quark chirality. The SU(2) relations of Eqs.~(\ref{su2releR})
and (\ref{su2relnu}) are imposed
and only one contact term at a time is assumed to be different from zero.
Combinations of contact interactions which are parity conserving or
blind to atomic parity violation constraints are
considered in Tables~\ref{table-new2} and \ref{table-new3}.

\begin{table}[h]
\caption[]{\label{table6}
The best estimate of the contact interaction parameters
$\eta_{\alpha\beta}^{eq}$ and the corresponding $\chi^2_{\rm min}$ for
(c) all eight $\eta_{\alpha\beta}^{eq}$ being free, and (d)--(f) various
chirality combinations.  The $\chi^2$ for the Standard Model and
the ``eye-ball'' solution from our previous paper are shown in
(a) and (b), respectively.
The data sets included are: H1, ZEUS, APV, $e$-N, DY, and LEP.
The number of degree of freedom for the Standard Model is 164.
}
\medskip
\begin{tabular}{|l||l|c|}
\multicolumn{1}{|c||}{Chirality Combination} & Best estimate (unit=TeV$^{-2}$)
 &  $\chi^2_{\rm min}$ \\
\hline
(a) Standard Model        &         & 176.4/164 d.o.f. \\
\hline
(b) $\eta_{LR}^{eu}=\eta_{RL}^{eu}=1.4 {\rm TeV}^{-2}$,
          &                           & 187.1/163 d.o.f. \\
{}   others=0 & &\\
\hline
(c) All free &$\eta_{LL}^{eu}=0.16
               \stackrel{\scriptstyle +0.61}{\scriptstyle -0.48},
   \eta_{RR}^{eu}=-0.001\stackrel{\scriptstyle +0.63}{\scriptstyle -0.55}$,
               &167.2/156 d.o.f.
\\
 & $\eta_{LR}^{eu}=0.80\stackrel{\scriptstyle +0.46}{\scriptstyle -0.62}$,
 $\eta_{RL}^{eu}=0.19\stackrel{\scriptstyle +0.65}{\scriptstyle -0.71}$, & \\
 & $\eta_{LL}^{ed}=0.31\stackrel{\scriptstyle +0.92}{\scriptstyle -1.01}$,
 $\eta_{RR}^{ed}=0.62\stackrel{\scriptstyle +0.98}{\scriptstyle -1.09},$ & \\
 & $\eta_{LR}^{ed}=1.66\stackrel{\scriptstyle +1.18}{\scriptstyle -1.66}$,
   $\eta_{RL}^{ed}=1.95\stackrel{\scriptstyle +1.10}{\scriptstyle -1.63}$, & \\
\hline
(d) $\eta_{RL}^{eu}=\eta_{LR}^{eu}$, $\eta_{RL}^{ed}=\eta_{LR}^{ed}$,
    & $\eta_{RL}^{eu}=\eta_{LR}^{eu}=0.51\stackrel{\scriptstyle +0.26}
      {\scriptstyle -0.27},$ & 170.0/ 162 d.o.f.   \\
{}{}{}
 others=0 & $\eta_{RL}^{ed}=\eta_{LR}^{ed}=1.97\stackrel{\scriptstyle +0.50}
        {\scriptstyle -0.70},$ &                     \\
\hline
(e) $\eta_{VV}^{eu}, \eta_{VV}^{ed}$ & $\eta_{VV}^{eu}=0.47
  \stackrel{\scriptstyle +0.20}{\scriptstyle -0.23}$,
    $\eta_{VV}^{ed}=1.10\stackrel{\scriptstyle +0.28}{\scriptstyle -0.38}$
                                        & 172.5/162 d.o.f. \\
\hline
(f) $\eta_{AA}^{eu}, \eta_{AA}^{ed}$ &
   $\eta_{AA}^{eu}=-0.36\stackrel{\scriptstyle +0.17}{\scriptstyle -0.14}$,
   $\eta_{AA}^{ed}=-0.36\stackrel{\scriptstyle +0.44}{\scriptstyle -0.46}$
                                     & 173.0 /162 d.o.f. \\
\hline
(g) SU(12)
   &  $\eta_{LL}^{eu}=-\eta_{LR}^{eu}= -0.39
    \stackrel{\scriptstyle +0.68}{\scriptstyle -0.31}$,
                  & 172.7/160 d.o.f. \\
   &   $\eta_{RL}^{eu}=-\eta_{RR}^{eu}=  0.29
      \stackrel{\scriptstyle +0.48}{\scriptstyle -0.68}$, & \\
   &   $\eta_{LL}^{ed}=-\eta_{LR}^{ed}= -0.81
      \stackrel{\scriptstyle +1.23}{\scriptstyle -0.96}$, & \\
   &   $\eta_{RL}^{ed}=-\eta_{RR}^{ed}=  -0.11
      \stackrel{\scriptstyle +1.24}{\scriptstyle -0.92}$ & \\
\end{tabular}
\end{table}

\begin{table}[h]
\caption[]{\label{table6b}
Breakdown of the minimum $\chi^2$ by experiment for the models
in Table \ref{table6}.
}
\medskip
\begin{tabular}{|l||cccccc|}
                & H1 & ZEUS & APV & $e$N & CDF-DY & LEP \\ \cline{2-7}
\multicolumn{1}{|r||}{\# Data points}
                    & 32 & 81   &  1  &  4  &  32    &  14    \\
\hline
\hline
(a) SM  & 33.9 & 59.9 & 1.2 & 1.8 & 62.2 & 17.4 \\
\hline
(b) $\eta_{LR}^{eu}=\eta_{RL}^{eu}=1.4 {\rm TeV}^{-2}$,
                &   29.0 & 66.5 & 1.2 & 1.8 &  70.9 & 17.7 \\
{}   others=0 &&&&&&\\
\hline
(c) All free & 29.9 & 58.1 & 0.0 & 0.5 & 62.6 & 16.1  \\
\hline
(d) $\eta_{RL}^{eu}=\eta_{LR}^{eu}=0.51$,
    & 30.0 & 58.2 & 1.2 & 1.8 & 62.7 & 16.1  \\
{} $\eta_{RL}^{ed}=\eta_{LR}^{ed}=1.97$ &&&&&& \\
\hline
(e) $\eta_{VV}^{eu}=0.47,\eta_{VV}^{ed}=1.10$
     & 30.4 & 58.7 & 1.2 & 1.8 & 63.1 & 17.4  \\
\hline
(f) $\eta_{AA}^{eu}=-0.36, \eta_{AA}^{ed}=-0.36$
     & 31.9 & 58.8 & 1.2 & 1.8 & 62.2 & 17.1 \\
\hline
(g) $\eta_{LL}^{eu}=-\eta_{LR}^{eu}= -0.39$,
   & 32.0 & 58.8 & 1.2 & 1.5 & 62.3 & 17.0 \\
{}   $\eta_{RL}^{eu}=-\eta_{RR}^{eu}=  0.29$, &&&&&& \\
{}   $\eta_{LL}^{ed}=-\eta_{LR}^{ed}=-0.81$, &&&&&&\\
{}   $\eta_{RL}^{ed}=-\eta_{RR}^{ed}=-0.11 $ & & & & & & \\
\end{tabular}
\end{table}

\def\err#1#2{$\stackrel{\scriptstyle +#1}{\scriptstyle -#2}$}

\begin{table}[h]
\caption[]{\label{table6c}
The best estimate of the $\eta_{\alpha\beta}^{eq}$ parameters when various
data sets are added successively.  In the last column when the $\nu$-N
data are included the $\eta_{L\beta}^{\nu q}$ are given in terms of
$\eta_{L\beta}^{eq}$ by Eq.~(\protect\ref{su2relnu}),
we assume $\eta_{RL}^{eu}=\eta_{RL}^{ed}$ in the last column.
}
\medskip
\begin{tabular}{|l|c|c|c|c|c|c|}
 & HERA only & HERA+APV & HERA+APV & HERA+APV & HERA+APV  &  HERA+DY+APV \\
 &           &          & +eN      & +eN+DY   & +eN+DY+LEP &+eN+LEP+$\nu$N \\
\hline
$\eta_{LL}^{eu}$ & 0.38\err{4.10}{6.04}  & 0.69\err{3.65}{5.49}
  & 0.49\err{2.90}{2.99}  & 0.16\err{0.67}{0.53}
  & 0.16\err{0.61}{0.48}  & -0.082\err{0.67}{0.40} \\
$\eta_{LR}^{eu}$ & -4.93\err{2.81}{0.96} & -4.90\err{6.41}{9.72}
  & -2.24\err{4.06}{2.90}  & 0.88\err{0.43}{0.64}
 & 0.80\err{0.46}{0.62}  & 0.86\err{0.42}{0.73} \\
$\eta_{RL}^{eu}$ & -0.14\err{3.29}{5.07} & -0.39\err{3.52}{4.80}
  & -4.20\err{4.51}{1.08}  & 0.21\err{0.68}{0.78}
  & 0.19\err{0.65}{0.71}  & 0.38\err{0.67}{0.91} \\
$\eta_{RR}^{eu}$ & 1.28\err{4.42}{6.31}  & 1.03\err{4.45}{5.90}
 & 1.46\err{1.73}{3.73} & 0.011\err{0.65}{0.61}
 & -0.0010\err{0.63}{0.55} & -0.074\err{0.74}{0.57} \\
$\eta_{LL}^{ed}$ & -2.51\err{10.49}{8.24} & -1.88\err{8.97}{7.59}
 & -4.03\err{7.45}{4.09}  & 0.71\err{1.38}{1.59}
 & 0.31\err{0.92}{1.01}  & 0.015\err{0.88}{0.52} \\
$\eta_{LR}^{ed}$ & -1.42\err{7.40}{4.87} & -1.19\err{7.04}{4.95}
 & -1.97\err{5.87}{3.77}  & 1.15\err{1.38}{1.99}
 & 1.66\err{1.18}{1.66}  & 0.88\err{1.06}{1.82} \\
$\eta_{RL}^{ed}$ & -2.80\err{7.39}{5.16} & -3.34\err{7.23}{4.31}
 & -2.80\err{5.99}{2.65}  & 1.50\err{1.29}{1.97}
 & 1.95\err{1.10}{1.63}  & $=\eta_{RL}^{eu}$ \\
$\eta_{RR}^{ed}$ & -3.16\err{10.85}{9.18} & -4.13\err{9.20}{6.43}
 & -2.40\err{7.63}{5.48}  & 1.01\err{1.31}{1.57}
 & 0.62\err{0.98}{1.09}  & 0.84\err{0.85}{1.11} \\
\hline
\hline
HERA    & 83.4 & 83.4 & 83.6  & 88.1  &  88.0  & 88.6  \\
APV	&      & 0.0  & 0.0   & 0.001 &  0.0   & 0.001  \\
eN      &      &      & 0.38  & 0.47  &  0.47  & 0.67  \\
DY      &      &      &       & 62.3  &  62.6  & 62.2 \\
LEP     &      &      &       &       &  16.1  & 17.1 \\
$\nu$N  &      &      &       &       &        &  0.007   \\
\hline
\hline
Total $\chi^2$ & 83.4 & 83.4 & 84.0  & 150.8 &  167.2 & 168.7  \\
\hline
SM $\chi^2$ & 93.8 & 95.0 & 96.8 & 159.0 & 176.4 & 176.4 \\
\hline
SM d.o.f.   & 113    & 114    & 118     & 150     & 164     & 165
\end{tabular}
\end{table}

\begin{table}[h]
\caption[]{\label{table-new1}
The best estimate on $\eta_{\alpha\beta}^{eq}$ and the 95\% CL limits on the
contact interaction scale $\Lambda_{\alpha\beta}^{eq}$,
where $\eta_{\alpha\beta}^{eq}=4\pi\epsilon/(
\Lambda_{\alpha\beta\epsilon}^{eq})^2$.
When one of the $\eta$'s is considered the others are set to zero.
SU(2) relations are assumed and $\nu$N data are included.
}
\medskip
\begin{tabular}{|cc|cc|}
\hline
& & \multicolumn{2}{c|}{95\% CL Limits} \\
Chirality  ($q$) &  $\eta\;\;$ (TeV$^{-2}$)  & $\Lambda_+$ (TeV)
                                             & $\Lambda_-$ (TeV) \\
\hline
\hline
LL($u$)  & $0.046 \pm 0.057$ &  9.3  & 12.0 \\
LR($u$)  & $0.11  \pm 0.079$ &  7.2  & 11.3 \\
RL($u$)  & $-0.041\pm 0.038$ &  15.4 & 10.9 \\
RR($u$)  & $-0.11 \pm 0.080$ &  11.1 & 7.3 \\
LL($d$)  & $0.055 \pm 0.061$ &  8.8  & 11.9 \\
LR($d$)  & $0.076 \pm 0.072$ &  7.9  & 11.2\\
RR($d$)  & $-0.074\pm 0.072$ &  11.2 & 8.0
\end{tabular}
\end{table}

\begin{table}[h]
\caption[]{\label{table-new2}
The best estimate on $\eta^{eq}$ for the minimal setting, $VV,AA$, and
SU(12), and the corresponding 95\% CL limits on the
contact interaction scale $\Lambda$, where $\eta=4\pi\epsilon/
(\Lambda_{\epsilon})^2$.
When one of the $\eta$'s is considered the others are set to zero.
Here we do not use SU(2) relations nor do we include the $\nu$N data.
}
\medskip
\begin{tabular}{|cc|cc|}
\hline
& & \multicolumn{2}{c|}{95\% CL Limits} \\
Chirality  ($q$) &  $\eta\;\;$ (TeV$^{-2}$)  & $\Lambda_+$ (TeV)
                                             & $\Lambda_-$ (TeV) \\
\hline
\hline
$\eta_{LR}^{eu}=\eta_{RL}^{eu}$  & 0.58 \err{0.27}{0.32}   & 3.6  & 5.8 \\
$\eta_{LR}^{ed}=\eta_{RL}^{ed}$  & $2.07$  \err{0.46}{0.63} & 2.1  & 2.6  \\
\hline
$\eta_{VV}^{eu}$   & 0.024\err{0.15}{0.13} &  6.2 & 7.5 \\
$\eta_{VV}^{ed}$   & 0.29\err{0.38}{0.57}  &  3.8 & 2.6 \\
\hline
$\eta_{AA}^{eu}$  & $-0.28$\err{0.16}{0.14} & 6.5  &  5.1 \\
$\eta_{AA}^{eu}$  & 0.21 \err{0.29}{0.35} & 4.3  &  5.1 \\
\hline
$\eta_{LL}^{eu}=-\eta_{LR}^{eu}$  & $-0.42$\err{0.26}{0.21} & 4.2 & 4.1 \\
$\eta_{RL}^{eu}=-\eta_{RR}^{eu}$  & 0.51\err{0.21}{0.26} & 3.9 & 3.9 \\
$\eta_{LL}^{ed}=-\eta_{LR}^{ed}$  & 0.31\err{0.50}{0.70} & 3.4 & 2.2 \\
$\eta_{RL}^{ed}=-\eta_{RR}^{ed}$  & $-0.51$\err{0.70}{0.50} & 2.3 & 3.1
\end{tabular}
\end{table}

\begin{table}[h]
\caption[]{\label{table-new3}
Same as Table II but with a further condition: $\eta^{eu}=\eta^{ed}$.
Here $q=u=d$.
}
\medskip
\begin{tabular}{|cc|cc|}
\hline
& & \multicolumn{2}{c|}{95\% CL Limits} \\
Chirality  ($q$) &  $\eta\;\;$ (TeV$^{-2}$)  & $\Lambda_+$ (TeV)
                                             & $\Lambda_-$ (TeV) \\
\hline
\hline
$\eta_{LR}^{eq}=\eta_{RL}^{eq}$  & 0.69 \err{0.23}{0.30} & 3.5  & 5.3 \\
\hline
$\eta_{VV}^{eq}$   & 0.11\err{0.34}{0.21}  &  4.1  & 6.9 \\
\hline
$\eta_{AA}^{eq}$  & $-0.36$\err{0.17}{0.14} & 4.4   & 4.7 \\
\hline
$\eta_{LL}^{eq}=-\eta_{LR}^{eq}$  & $-0.54$\err{0.26}{0.20} & 3.5 & 3.8 \\
$\eta_{RL}^{eq}=-\eta_{RR}^{eq}$  & 0.58\err{0.20}{0.26}  & 3.8 & 3.4
\end{tabular}
\end{table}

\section{Conclusions}

In summary, we have made a general investigation to determine whether
the excess of events observed at high-$Q^2$ in the HERA experiments can
be understood in terms of contact interactions.
All available low and high energy data relevant to $eeqq$ contact terms
have been included in a global fit. Our findings can be summarized as follows:

\noindent
(i) If the HERA high $Q^2$ events are due to $eeqq$ contact interactions,
$\eta_{LR}^{eu}$ and $\eta_{RL}^{eu}$ terms are the most effective in
explaining the data.

\noindent
(ii) The CDF data on high mass Drell-Yan lepton-pair production, however,
strongly restrict the $eeuu$ contact interactions irrespective of their
chirality structure.

\noindent
(iii) Once the high-energy data and the low-energy electroweak data are
combined, all $eeqq$ contact interactions are strongly constrained and
the possibility of explaining the HERA high $Q^2$ events with contact
interactions is limited.

\noindent
(iv) The SM gives a reasonably good overall description of all the
electron-hadron data, including the HERA high $Q^2$ events.

\noindent
(v) The limits obtained on the contact scales are more restrictive than 
those obtained previously~\cite{pdg}.

\section*{Acknowledgments}
K.C.\ would like to thank S.~Godfrey and K.~McFarland for discussions.
We thank Dieter Haidt for discussions of error correlations in the HERA data.
This research was supported in part by the JSPS-NSF Joint Research Project,
in part by the U.S.~Department of Energy under Grant
Nos. DE-FG03-93ER40757 and DE-FG02-95ER40896 and in part
by the University of Wisconsin Research Committee with funds granted
by the Wisconsin Alumni Research Foundation.

\end{document}